\documentclass[pra,aps,graphics,a4paper,10pt,twocolumn,accepted=2021-03-08]{quantumarticle}
\pdfoutput=1
\usepackage[margin=25mm]{geometry}
\usepackage[export]{adjustbox}
\usepackage{amsmath,bbm}
\usepackage{braket}
\usepackage{ragged2e}
\usepackage{stackengine}

\usepackage{tikz}
\usepackage{pgfplots}
\usepackage{tikzscale}
\usepackage{siunitx}

\usepackage{url}

\usepackage{breakurl}
\usepackage[breaklinks]{hyperref}

\usepackage{graphicx}
\usepackage[]{subcaption}

\newcommand{\changes}[1]{{\color[rgb]{0, 0, 0}{#1}}}

\newcommand{\be}{\begin{equation}}
\newcommand{\ee}{\end{equation}}
\newcommand{\ba}{\begin{array}}
\newcommand{\ea}{\end{array}}
\newcommand{\bqa}{\begin{eqnarray}}
\newcommand{\eqa}{\end{eqnarray}}

\newcommand{\um}{\mathbbm{1}}

\newcommand{\prj}[1]{\ensuremath{| #1 \rangle \langle #1 |}}

\newcommand{\matel}[3]{\ensuremath{\langle #1 | #2 | #3 \rangle}}

\newcommand{\Inv}{{\cal I}}
\newcommand{\Itwo}{\Gamma}
\newcommand{\Ione}{\vec{W}}
\newcommand{\IO}{\theta}
\newcommand{\Htwo}{\Omega}
\newcommand{\Hone}{\vec{V}}
\newcommand{\Htrapfreq}{M}
\newcommand{\Hforce}{\vec{F}}
\newcommand{\Hcenter}{\vec{C}}
\newcommand{\Cov}{\Sigma}
\newcommand{\traj}{\vec{L}}
\newcommand{\symplectic}{{\cal S}}
\newcommand{\antisyminteg}{{\cal J}}
\newcommand{\pos}{R}
\newcommand{\com}{A}

\newcommand{\unitary}{U}
\newcommand{\ansatzparam}{P}
\newcommand{\linearinv}{\vec{Z}}

\newcommand{\duration}{T}

\begin{document}

\title{Quantum control with a multi-dimensional Gaussian quantum invariant}

\author{Selwyn Simsek}
\affiliation{Physics Department,	Blackett Laboratory, Imperial College London, Prince Consort Road, SW7 2BW, United Kingdom}
\author{Florian Mintert}
\affiliation{Physics Department,	Blackett Laboratory, Imperial College London, Prince Consort Road, SW7 2BW, United Kingdom}
\date{March 8, 2021}

\begin{abstract}
The framework of quantum invariants is an elegant generalization of adiabatic quantum control to control fields that do not need to change slowly.
Due to the unavailability of invariants for systems with more than one spatial dimension, the benefits of this framework have
not yet been exploited in multi-dimensional systems.
We construct a multi-dimensional Gaussian quantum invariant that permits the design of time-dependent potentials that let the ground state of an initial potential evolve towards the ground state of a final potential.
The scope of this framework is demonstrated with the task of shuttling an ion around a corner which is a paradigmatic control problem in achieving scalability of trapped ion quantum information technology.
\end{abstract}

\maketitle

\section{Introduction}

The development of hardware for quantum information processing has reached a stage in which tens of qubits can be accurately controlled \cite{friis_observation_2018,wei_verifying_2020}.
In order to control such a large register of qubits, it is advisable to arrange them into several smaller units that are coherently interconnected.
Such interconnections can be realised in a multitude of different ways \cite{monroe_large-scale_2014,Lekitsch2017}.
In the case of trapped ions, one of the most prominent envisioned architectures is based on segmented traps, in which ions can be moved, {\it i.e.} shuttled, between different segments \cite{Kielpinski2002,Lekitsch2017} of a segmented trap.

Since trapped ion quantum logic usually requires the motional states of ions to be
close to their quantum mechanical ground state \cite{molmer_multi-particle_1999},
it is desirable that any such shuttling process should transfer ions to their motional ground state with high fidelity.
Performing all operations of a quantum algorithm within the time-scale imposed by the system's coherence times, requires fast shuttling \cite{Kielpinski2002}
implying that the transient motional states during shuttling are far away from the motional ground state.
Any shuttling process thus needs to end with a stage of deceleration in which an originally rapidly moving ion is being transferred to its quantum mechanical motional ground state.

In practice, shuttling is realised in terms of time-dependent trapping potentials resulting from voltages applied to trap electrodes \cite{Rowe2002,Hensinger2006}.
Describing the quantum dynamics of a particle through such a time-dependent potential landscape is generally a formidable numerical challenge,
but in the present context the problem can be substantially simplified.
The initial ground state is a Gaussian wavepacket characterised by the expectation values of position and momentum and the corresponding covariances of those operators.
For realistic trapping potentials, the potential can be approximated to be harmonic in the spatial domain occupied by such a wavepacket \cite{leibfried_quantum_2003}.
Within this approximation, an initial Gaussian wavepacket remains Gaussian under time evolution, and thus the state is characterised in terms of expected position in phase space and covariances, {\it i.e.} a small number of parameters \cite{heller_timedependent_1975,Heiler1975,heller_classical_1976,heller_generalized_1977,heller_frozen_1981}.

Identifying time-dependent gate voltages that realise a shuttling protocol that transfers one or several ions from their ground state in the initial potential to the ground state of their final potential, typically requires testing a large number of different potential solutions.
As such, the reduction of numerical effort resulting from the Gaussian approximation is invaluable.

An equally important reduction in effort can be achieved with the framework of quantum invariants \cite{lewis_motion_1968},
since those allow one to directly construct shuttling protocols that end with the ions in their motional ground state \cite{chen_lewis-riesenfeld_2011}.
The framework of quantum invariants and its use for control of wave-packets via
invariant-based inverse engineering is well-established for one-dimensional problems \cite{chen_lewis-riesenfeld_2011,torrontegui_fast_2011,levy_noise_2018,torrontegui_chapter_2013-1,tobalina_fast_2017}.

Invariant-based inverse engineering has been also suggested to realise the transport of \changes{cold atoms in optical lattices~\cite{e22030262}
or optical tweezers~\cite{zhang_optimal_2016}} as well as atom clouds with a corresponding experimental implementation \cite{ness_realistic_2018}. Invariant-based inverse engineering has also motivated the study of fast transport of spin-orbit-coupled Bose-Einstein condensates \cite{chen_inverse_2018}, and quantum invariants have been used as a theoretical tool to compute topological phases in planar waveguides \cite{rosu_ermakov_1999} and minispace quantum cosmologies \cite{espinoza_ermakov-lewis_2000}. They may also be constructed for light beam propagation in nonlinear inhomogeneous media \cite{goncharenko_ermakov_1991}.

Processes in which an ion is being shuttled across a junction in a more complicated trap geometry, however, cannot yet be approached with invariant-based inverse engineering,
\changes{because the development of invariants for systems with more than one translational degree of freedom \cite{PhysRevA.102.063112}is still in its infancy.}
Here, we develop a framework of quantum invariants that is suitable for optimal control of Gaussian wave-packets in any number of spatial dimensions via invariant-based inverse engineering. 

\changes{In Sec.~\ref{sec:invariant-control}, invariant-based inverse engineering and Gaussian states are summarised. Sec.~\ref{sec:optimal-control} introduces the invariant that is at the core of this paper and discusses how this invariant can be used to design time-dependent potentials that result in ground-state-to-ground-state shuttling.
This section is followed by explicit examples of such protocols discussed in Sec.~\ref{sec:protocols},
while the actual derivation of the invariant is in Sec.~\ref{sec:invariant}, and Sec.~\ref{sec:one-dimensional} contains a discussion of how well-established one-dimensional invariants are contained as special case in the present framework.
A reader who is more interested in the underlying framework than in its application to a specific control problem, can skip Sec.~\ref{sec:protocols} or read Secs.~\ref{sec:invariant} and \ref{sec:one-dimensional} in any order.}

\section{Invariant-based control}
\label{sec:invariant-control}

In this section several elements of Gaussian quantum states and quantum invariants upon which the present work builds will be reviewed.

In order to ease notation, from hereon in the momentum variables $p_i$ will include a scaling factor depending on the mass of each particle, such that the kinetic energy reads
$\sum_i\hat p_i^2/(2m)$ instead of the usual expression $\sum_i\hat p_i^2/(2m_i)$ that explicitly reflects the possibility of having different ions with different masses.
A corresponding scaling of the position variables $x_i$ ensures that $\hat x_i$ and $\hat p_i$, as well as their classical counterparts, are conjugate to each other.

\subsection{Gaussian dynamics}

The framework uses the Gaussian approximation of wavepackets that simplifies the description in terms of general quantum states to a description in terms of classical trajectories and covariances.
If $\hat X$ denotes the $2d$-dimensional vector comprised of the phase space operators of a $d$-dimensional system, {\it i.e.} $d$ position operators and $d$ momentum operators
\be
\hat X =  \begin{pmatrix} \hat r_1 , \hat r_2 , ... , \hat r_n , \hat p_1 , \hat p_2 , ... , \hat p_n\end{pmatrix},
\ee
then any Gaussian wavepacket is completely characterised in terms of the expectation value of $\hat X$, denoted in the following by the vector $X$,
and by the $2d\times 2d\ -$ dimensional covariance matrix $\Cov$ with the elements
\be
\Cov_{ij}=\frac{1}{2}\langle \hat X_i\hat X_j+\hat X_j\hat X_i\rangle-\langle \hat X_i\rangle\langle\hat X_j\rangle\ ,
\ee
where the symbol $\langle\ \rangle$ denotes the expectation value with respect to the Gaussian quantum state.

In a potential that is not always harmonic, an initially Gaussian quantum state will lose its Gaussian character,
but if the wavepacket is sufficiently well localised in real-space, then the potential can be approximated by its second-order Taylor expansion around the center of the wavepacket, and the quantum state remains Gaussian~\cite{Heiler1975}.

The center of the wavepacket, $X=\langle \hat X \rangle$, follows the classical trajectory, {\it i.e.} it satisfies the equation of motion
\be 
\dot X= \symplectic \Hone\ ,
\label{eq:F}
\ee
where
\be
\Hone=\nabla H_c
\ee
is the vector containing the first derivatives of the {\em classical} Hamiltonian $H_c$ with respect to the phase space variables, evaluated at the centre of the wavepacket $X$, and the symplectic matrix
\be
\symplectic=\left[
  \ba{rr}
  \mathbbm{O} & \um \\
  -\um & \mathbbm{O} \\
  \ea\right]\ ,
\ee
is defined in terms of the $d\times d-$dimensional identity matrix $\um$ and zero-matrix $\mathbbm{O}$.

The covariance matrix satisfies the equation of motion \cite{Mintert2009}
\be
\dot\Cov= \Cov \Htwo \symplectic - \symplectic \Htwo \Cov \\
\ee
in terms of the matrix $\Htwo$ containing the second derivatives
\be
\Htwo_{ij} = \frac{ \partial ^2 H_c}{\partial x_i \partial x_j}
\label{eq:Omega}
\ee
of the classical Hamiltonian $H_c$ with respect to the phase space variables evaluated at $X$.

From hereon in it is assumed that the anharmonic component of $H$, which is to say the Taylor expansion of $H$ beyond the second order, is negligible, in which case the approximation holds and the system Hamiltonian can be expressed as
\be
H= \frac{1}{2} \hat X ^T \Htwo \hat X + \Hone\hat X\ .
\ee

\subsection{Quantum invariants}
\label{sec:invariants}

The problem at hand lies in finding $\Htwo$ and $\Hone$ such that $X(\duration)$ matches the desired phase space position at the final instance in time $\duration$ of the shuttling protocol, and such that $\Cov(\duration)$ corresponds to the covariances of the quantum mechanical ground state of $H(\duration)$.
Quantum invariants allow one to derive straightforwardly expressions for $\Htwo$ and $\Hone$ that ensure that the boundary conditions at $t=0$ and $t=\duration$ are satisfied \cite{tobalina_fast_2017}.

A quantum invariant is any operator $\Inv(t)$ satisfying the equation of motion \cite{lewis_classical_1967}
\be \label{eqn:invariant-eom}
\frac{\partial \Inv(t)}{\partial t}=i[\Inv(t),H(t)]\ .
\ee

Crucially, the instantaneous eigenstates of an invariant, are solutions of the Schr\"odinger equation with the time-dependent Hamiltonian $H(t)$ \cite{lewis_motion_1968}.
A quantum invariant with non-degenerate spectrum that commutes with the Hamiltonian at the beginning and the end of a time interval of interest,
{\it i.e.} $[\Inv(0),H(0)]=0$ and $[\Inv(\duration),H(\duration)]=0$, can thus be used to identify shuttling protocols that ensure that the system ends up in an eigenstate of the final Hamiltonian $H(\duration)$, if it is initialised in an eigenstate of $H(0)$ \cite{chen_lewis-riesenfeld_2011}.
If the system is initialised in the ground state of $H(0)$, then it will be transferred to the ground state of $H(\duration)$ since the only eigenstate of the quadratic operator $H(\duration)$ that is Gaussian is the ground state. This is similar to the idea of adiabatic transitions, but in contrast, the invariant based method is exact and holds also under fast changes in the Hamiltonian,
and the system state remains an eigenstate of the invariant $\Inv(t)$ rather than an eigenstate of the system Hamiltonian $H(t)$.

\subsection{Control with quadratic invariants}

In the case of quadratic, time-dependent, Hamiltonians,
quantum invariants can be constructed from which one 
can deduce quadratic Hamiltonians $H(t)$ that correspond to those invariants.
This, however, is not very useful in practice since,
the resulting Hamiltonians contain in general bilinear terms of phase space operators;
in particular, the part corresponding to the kinetic energy will be of the form $\sum_{ij}h_{ij}(t)\hat p_i\hat p_j$, which is not necessarily equal to the desired term $\sum_i\hat p_i^2/2m$.

Since, in practice, shuttling must be realised in terms of modulation of only the trapping potential,
it is crucial to ensure that Hamiltonians resulting from the invariant-based framework are of the form
\be
H(t)=\sum_i\frac{\hat p_i^2}{2m}+\frac{1}{2}m\sum_{ij}\Htrapfreq_{ij}\hat x_i\hat x_j-\sum_i\Hforce_i\hat x_i\ ,
\label{eq:HTV}
\ee
with tuneable parameters $\Hforce$ and $\Htrapfreq$,
and no Lorentz-type terms of the form $\hat x_i\hat p_j+\hat x_j\hat p_i$, or terms linear in the momenta $\hat p_i$.

In one-dimensional systems, there does exist a class of quantum invariants \cite{chen_lewis-riesenfeld_2011} that ensures the Hamiltonians are of this form and gives a way to deduce $\Htrapfreq$ and $\Hforce$ from the invariant. Despite the utility and applicability of these one-dimensional invariants, generalisations to higher dimensional systems have not been found to date.

\section{Optimal control with high-dimensional quadratic invariants}
\label{sec:optimal-control}

As the present work deals with quadratic Hamiltonians, it is most useful to work with an invariant that contains the phase space operators $\hat X$ only up to second order.
Any such invariant can be written as
\be
\Inv(t)=\frac{1}{2}\hat X^T\Itwo\hat X+\Ione \cdot \hat X+\IO\ ,
\label{eq:Inv}
\ee
parametrised in terms of a time-dependent, Hermitian matrix $\Itwo(t)$, a real vector $\Ione(t)$, and a real scalar $\IO(t)$.

$\Inv(t)$ is an invariant if and only if Eq.~\eqref{eqn:invariant-eom} holds, which is the case if the equations
\begin{align}
\frac{d\Itwo}{d t}&=\Htwo \symplectic \Itwo-\Itwo \symplectic \Htwo ,\label{eq:eominvtwo}\\
\frac{d\Ione}{d t}&=\Htwo \symplectic \Ione-\Itwo \symplectic \Hone ,\label{eq:eominvone}\\
\frac{d\IO}{d t}&=\Hone^T \symplectic \Ione ,\label{eq:eominvo}
\end{align}
with $\Htwo$ and $\Hone$ as defined in Eqs.\eqref{eq:F} and \eqref{eq:Omega} are satisfied.
For any choice of $\Itwo$ and $\Ione$,
one may deduce choices of $\Htwo$, $\Hone$ and $\IO$ which satisfy Eqs.~\eqref{eq:eominvtwo}, \eqref{eq:eominvone} and \eqref{eq:eominvo}.

This approach would permit us to find a quadratic Hamiltonian for an invariant specified by functions $\Itwo(t)$ and $\Ione(t)$ that one would be free to choose.
For most choices of such invariants, however, any compatible Hamiltonian will not necessarily be of the form given in Eq.~\eqref{eq:HTV},
{\it i.e.} it would not be realizable in practice.
It is thus essential to find instances of $\Itwo$ and $\Ione$, such that $\Htwo$ is of the form
\be
\Htwo=\left[
\ba{cc}
m \Htrapfreq & \mathbbm{O}\\
\mathbbm{O} & \frac{1}{m}\um
\ea\right]\ ,
\label{eq:Wansatz}
\ee
and such that  $\Hone$ is of the form
\be
\Hone=\left[\ba{c}-\Hforce \\ 0\ea \right]\ .
\label{eqn:ansatz-v}
\ee

\subsection{A suitable invariant}
\label{sec:def:invariant}

While a general invariant of the form specified in Eq.~\eqref{eq:Inv} does not correspond to a Hamiltonian of desired form,
the following explicit parametrization forces the Hamiltonian to be of the form of Eq.~\eqref{eq:HTV}, as shown in Sec.\ref{sec:invariant}.

The quadratic part $\Gamma$ of the invariant is given by
\be
\ba{l}
\Itwo=\\
\left[
\ba{cc}
\displaystyle m \left( \dot{\pos}^2 +  \Re\left([\dot{\pos},\pos]_{\com} - \pos \com^2 \pos\right) \right)&  
\displaystyle \frac{\antisyminteg-\{\pos,\dot{\pos}\}}{2}\\[6pt]
\displaystyle \frac{- \antisyminteg -\{\pos,\dot{\pos}\}}{2}&
\displaystyle\frac{\pos^2}{m} 
\ea \right]
\ea
\label{eqn:ansatz-gamma} \nonumber
\ee
where the symbol 
$[x,y]_{z}=xzy-yzx$ denotes a generalized commutator and $\{x,y\}=xy+yx$ is the anti-commutator.

$\Gamma$ is parametrised in terms of a $d\times d$ real, positive and semi-definite square matrix $\pos$ satisfying $\dot \pos(0)=0$,
and the matrices $\com$ and $\antisyminteg$ are determined uniquely from the relations
\bqa
\com&=&
i \pos^{-2} + \frac{1}{2} [ \pos^{-1}, \dot{\pos}] +
 \frac{1}{2} \pos^{-1} {\antisyminteg} \pos^{-1}\ , \nonumber \\
\{ {\antisyminteg} , \pos^{-2} \} &=& [\dot \pos, \pos^{-1}] + [ \pos, \pos^{-2}]_{\dot\pos}\ . \label{eqn:first-relations}
\eqa

The linear part $\Ione$ of the invariant $\Inv$ is given by
\be
\Ione=-\Itwo\left[\ba{c}\traj\\m \dot \traj\ea\right]\ , \label{eqn:ansatz-w}
\ee
parametrized by a d-dimensional real vector $\traj$,

and the scalar part is given by

\be
\IO(t)=\frac{1}{2}\Ione^T\Itwo^{-1}\Ione\ . \label{eqn:ansatz-scalar}
\ee

\changes{The Hamiltonian corresponding to an invariant is determined by the equations of motion Eqs.~\eqref{eq:eominvtwo} and \eqref{eq:eominvone}.
As shown in the following,
the quantities $\Htrapfreq$ and $\Hforce$ (defined in Eq.~\eqref{eq:Wansatz} and \eqref{eqn:ansatz-v}) used to parametrize and accessible Hamiltonians
are determined by Eqs.~\eqref{eqn:first-relation-1} and \eqref{eqn:first-relation-2}.

The quadratic component $\Htrapfreq$ is determined by
\be
\{\pos^2,\Htrapfreq\}=2[\dot\pos,\pos]_{\com}-\{\ddot \pos, \pos\}-2\pos\com^2\pos \label{eqn:first-relation-1}
\ee
in terms of $R$ following Eq.~\eqref{eq:eomR}.
This is still not an explicit solution for $\Htrapfreq$, but it can be obtained in terms of an expansion into the eigenstates of $R$ as stated in Eq.~\eqref{eq:Omegax}.

Once $\Htrapfreq$ is obtained, the linear component $\Hforce$ is obtained through the relation
\be
\Hforce=m\left(\ddot \traj + \Htrapfreq \traj \right)\, \label{eqn:first-relation-2}
\ee
following Eq.~\eqref{eqn:vector-newton-equation}.

$\pos$ and $\traj$ may be chosen freely subject to a set of boundary conditions discussed in the next section.}

\subsection{Boundary conditions}

As stated in Sec.~\ref{sec:invariants}, a prerequisite for invariant-based control is that the invariant and Hamiltonian commute at initial and final times, {\it i.e.}
\be
[\Inv(0),H(0)]=[\Inv(\duration),H(\duration)]=0\ .
\label{eq:bundary}
\ee
Since the initial and final Hamiltonians $H(0)$ and $H(\duration)$ are determined by the choice of problem at hand,
this commutativity must be ensured only by imposing boundary conditions on the invariant $\Inv(t)$, and not $H(t)$.
Since the invariant $\Inv(t)$ is parametrised in terms of $\traj$ and $\pos$ via Eqs.~\eqref{eq:Inv}, \eqref{eqn:ansatz-gamma} and \eqref{eqn:ansatz-w},
this implies boundary conditions for $\traj$ and $\pos$.

The choices 
\be
\ba{rclcrcl}
\pos(t)&=&\Htrapfreq(t)^{-\frac{1}{4}}\ ,&&
\dot \pos(t)&=&0\ ,\\
\traj(t)&=&
(m\Htrapfreq(t))^{-1}\Hforce(t)\ ,
& \mbox{and} &\dot \traj(t)&=&0
\ea
\label{eq:boundary}
\ee
directly result in commutativity of $\Inv(t)$ and $H(t)$.
Imposing the conditions of Eq.~\eqref{eq:boundary} for $t=0$ and $t=\duration$ thus imposes the desired boundary conditions Eq.~\eqref{eq:bundary}.

An additional set of conditions arise from the requirement that $\Inv$ be an invariant. The equations
\bqa
\{\ddot \pos, \pos\}+\{\pos^2,\Htrapfreq\} &=& 2[\dot\pos,\pos]_{\com}-2\pos\com^2\pos \nonumber \\
\ddot \traj + \Htrapfreq \traj &=& \frac{\Hforce}{m},
\eqa
which are derived in Sec.~\ref{sec:invariant} as Eqs.~\eqref{eq:eomR} and \eqref{eqn:vector-newton-equation}, must be satisfied at all times. Substituting Eqs.~\eqref{eq:boundary} into these equations gives an additional set of boundary conditions
\be
\ddot \traj(t)=0\ ,\ \mbox{ and }\ \ddot \pos(t)=0.
\label{eq:moreboundary}
\ee

For any time other than $t=0$ and $t=\duration$, the conditions of Eqs.\eqref{eq:boundary} and \eqref{eq:moreboundary} do not need to be satisfied, and any time-dependent form of $\traj$ and $\pos$ that satisfies the boundary conditions may be employed.

In practice, however, some choices of $\traj(t)$ and $\pos(t)$ can result in trajectories $X(t)$ or covariances  $\Cov(t)$ of the quantum state that are conflicting with practical requirements, such as a trapped ion colliding with a trap electrode.
It is thus desirable to relate the variables $\traj(t)$ and $\pos(t)$ characterizing the invariant $\Inv$ to the variables $X(t)$ and  $\Cov(t)$ characterizing the quantum state.

Since the invariant $\Inv(t)$ is quadratic, this is most elegantly done via the fact that the time-dependent quantum state is the instantaneous ground state of $\Inv(t)$,
which is characterized by covariance matrix
\be
\Cov_g= \frac{1}{2}\Itwo^{-1}
\ee
and phase-space coordinate
\be
X_g= - \Itwo^{-1} \Ione\ .
\label{eq:sol:Ione}
\ee

Since $\Itwo$ (given in Eq.~\eqref{eqn:ansatz-gamma}) is symplectic, its inverse can be expressed as $\Itwo^{-1}=- \symplectic \Itwo \symplectic$, resulting in the explicit form
\be
  \label{eqn:cov-block}
 \ba{l}
  \Cov_g=\\
  \displaystyle\frac{1}{2} \left[\ba{cc}
    \displaystyle\frac{\pos^2}{m} &
\displaystyle\frac{\{\pos,\dot{\pos}\} - {\antisyminteg}}{2}\\[6pt]
\displaystyle\frac{\{\pos,\dot{\pos}\} + {\antisyminteg}}{2} &
m \left( \dot{\pos}^2 + \Re ([\dot{\pos},\pos]_{\com} - \pos \com^2 \pos) \right)
    \ea \right]
    \ea
\nonumber
\ee
for the covariance matrix.

With the form of $\Ione$ given in Eq.~\eqref{eqn:ansatz-w}, the explicit solution (Eq.~\eqref{eq:sol:Ione}) for $X_g$ reads
\be
X_g=\left[\ba{c}\traj\\m \dot \traj\ea\right]\ .
\label{eq:trajectory}
\ee
That is, $X_g$ is determined straightforwardly in terms of $\traj$,
which means that the trajectory can be easily chosen at will.

\changes{It is of course desirable to attribute an equally clear physical interpretation to the quantity $\pos$ as found for $\traj$ in the sense of the trajectory.
This, however, seems feasible only in the adiabatic limit, in which the time-derivative $\dot\pos$ becomes negligible.
$\antisyminteg$ in Eq.~\eqref{eqn:first-relations} thus also becomes negligible, and Eq.~\eqref{eqn:first-relation-1} reduces to
 \be
  \pos=\Htrapfreq^{-\frac{1}{4}}\ .
  \ee
  $\pos$ thus directly determines the curvature of the trapping potential.
  Apart from this limiting case, however,
  the relation between $\pos$ and the potential in the system Hamiltonian is substantially more complicated,
  consistent with the fact that diabatic dynamics is generally more complicated than adiabatic dynamics.
}

\section{Shuttling protocol}
\label{sec:protocols}

With the general framework for optimal control based on high-dimensional invariants established,
its practical use can be demonstrated with the problem of shuttling an ion through a two-dimensional potential landscape,
\changes{such as} the transfer from an initial trapping potential
\be
V_i=\frac{1}{2}m\left(\omega_t^{\changes{2}}x^2+\omega_r^{\changes{2}}(y-r)^2\right)
\label{eq:aux1}
\ee
centered around the initial position $[0,r]$
to a final potential 
\be
V_f=\frac{1}{2}m\left(\omega_r^{\changes{2}}(x-r)^2+\omega_t^{\changes{2}}y^2\right)
\label{eq:aux2}
\ee
centered around the final position $[r,0]$.
In addition to the change in position, in particular, also the shape of the potential will change during the shuttling process with the (typically strong) radial confinement \changes{(characterized by the frequency $\omega_r$) initially in the $y$-direction and finally in the $x$-direction,
and the (typically weak) axial confinement (characterized by the frequency $\omega_t$) initially in the $x$-direction and finally in the $y$-direction.}

As discussed above in the context of Eq.~\eqref{eq:trajectory}, the trajectory of the ion can be chosen at will, as long as it satisfies the required boundary conditions.
We are thus free to chose an arc
\be
\traj(t)= r \left[
\ba{c}
\displaystyle\sin \left( \frac{\pi}{2} p(\tau) \right)\vspace{.2cm} \\
\displaystyle\cos \left( \frac{\pi}{2} p(\tau) \right)
\ea
\right]
\ee
with $\tau=t/\duration$, radius $r$, and a function $p(\tau)$ satisfying the boundary conditions $p(0)=0$ and $p(1)=1$.
Motivated by its successful use in one-dimensional problems~\cite{torrontegui_fast_2011,Palmero2013,lu_fast_2014} we will choose
\be
p(\tau)=10\tau^3 - 15\tau^4 + 6\tau^5\ ,
\ee
which ensures that the boundary conditions
\be
\dot \traj(0)=\dot \traj(T)=\ddot \traj(0)=\ddot \traj(T)=0
\ee
are satisfied.

For the quadratic component $\pos(t)$ of the invariant, the boundary conditions read
\be
\pos(0)=
 \left[
    \ba{cc}
    \sqrt{\omega_r} & 0 \\
    0 & \sqrt{\omega_t} \\
    \ea
    \right]=:\pos_i\ ,
\ee
and
\be
\pos(T)=
 \left[
    \ba{cc}
    \sqrt{\omega_t} & 0 \\
    0 & \sqrt{\omega_r} \\
    \ea
    \right]=:\pos_f\ ,
\ee
as well as $\dot R=0$ and $\ddot R=0$ at $t=0$ and $t=\duration$.

A suitable time-dependent matrix $\pos(t)$ that satisfies the boundary conditions can be defined with the ansatz
\be
\pos(t)=(1-p(\tau))\pos_i+p(\tau)\pos_f+\tau^3 \left( 1-\tau\right) ^3R_c\ ,
\ee
The matrix $R_c$ that will determine how the trapping potential rotates along the trajectory of the ion can still be chosen at will,
and the following examples are based on the choice
\be
R_c=(\omega_t \omega_r)^\frac{1}{4} 
 \left[
    \ba{cc}
    1 & 1 \\
    1 & 1 \\
    \ea\right]\ .
\ee

With the given specific choices for $\pos(t)$ and $\traj(t)$ one can solve
Eqs.\eqref{eq:eominvtwo} and \eqref{eq:eominvone} for 
$\Htwo$ and $\Hone$, which yields the desired objects $\Htrapfreq$ and $\Hforce$ via
Eqs.\eqref{eq:Wansatz} and \eqref{eqn:ansatz-v}.

The explicit time-dependent solution for $\Htrapfreq$ depends on the system parameters  $\omega_r$ and $\omega_t$ and on the duration $\duration$ of the shuttling protocol.
Due to scale-invariance, the shuttling protocol depends only on two independent parameters,
and in the following, we will consider the variation of $\frac{\omega_r}{\omega_t}$ and $\omega_t\duration$.

The explicit time-dependent solution for $\Hforce$ depends additionally also on the mass $m$ of the ion and the radius $r$ of the trajectory.
Discussing the solution of the time-dependent trapping potential in terms of its center \changes{$\Hcenter=\frac{1}{m}\Htrapfreq^{-1}\Hforce$} instead of the force $\Hforce$ yields a solution that is independent of $m$, and that simply scales linearly with $r$.

As such, one can exemplify shuttling protocols for all values of $\omega_r$, $\omega_t$, $m$, $r$ and $T$ in terms of the two independent parameters $\frac{\omega_r}{\omega_t}$ and $\omega_t\duration$.

\begin{figure*}[!htb]
  \begin{subfigure}[t]{0.3333\textwidth}
    \def\stackalignment{l}
  \bottominset{(a)}{\includegraphics[width=\linewidth,clip,trim=0.3cm 0.1cm 0.2cm 0.3cm]{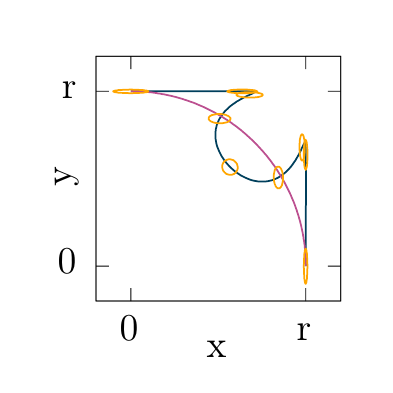}}{1.5cm}{1.3cm}
  \caption*{\changes{$\displaystyle\frac{\omega_r}{\omega_t} = 10\ $, $\ \omega_t T = 3$}}
 \end{subfigure}
  \begin{subfigure}[t]{0.3333\textwidth}
    
    \def\stackalignment{l}
  \bottominset{(b)}{\includegraphics[width=\linewidth, clip, trim=0.3cm 0.1cm 0.2cm 0.3cm]{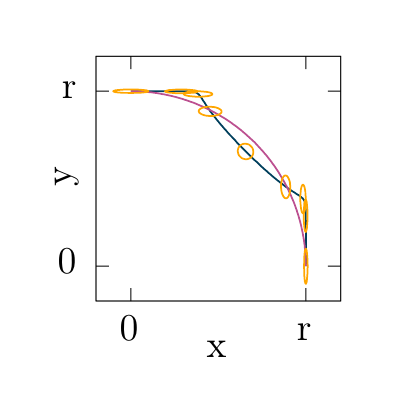}}{1.5cm}{1.3cm}
  \caption*{\changes{$\displaystyle\frac{\omega_r}{\omega_t} = 10\ $, $\omega_t T = 5$}}
\end{subfigure}
  \begin{subfigure}[t]{0.3333\textwidth}
    \def\stackalignment{l}
    \bottominset{(c)}{\includegraphics[width=\linewidth, clip, trim=0.3cm 0.1cm 0.2cm 0.3cm]{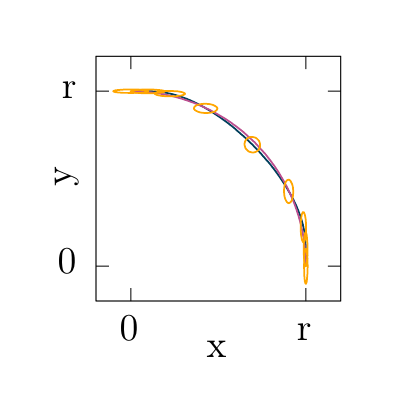}}{1.5cm}{1.3cm}
  \caption*{\changes{$\displaystyle\frac{\omega_r}{\omega_t} = 10\ $, $\omega_t T = 10$}}
\end{subfigure}
  \begin{subfigure}[t]{0.3333\textwidth}
    \def\stackalignment{l}
    \bottominset{(d)}{\includegraphics[width=\linewidth, clip, trim=0.3cm 0.1cm 0.2cm 0.3cm]{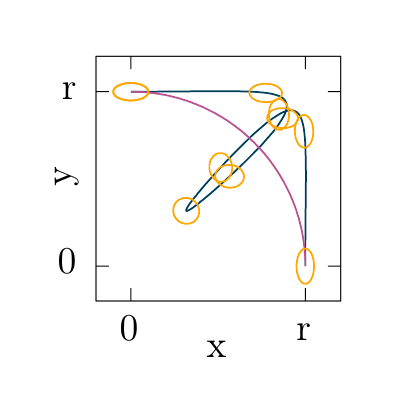}}{1.5cm}{1.3cm}
  \caption*{\changes{$\displaystyle\frac{\omega_r}{\omega_t} = 2\ $, $\omega_t T = 3$}}
\end{subfigure}
  \begin{subfigure}[t]{0.3333\textwidth}
    \def\stackalignment{l}
    \bottominset{(e)}{\includegraphics[width=\linewidth, clip, trim=0.3cm 0.1cm 0.2cm 0.3cm]{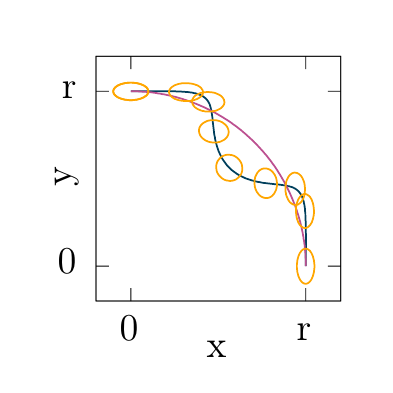}}{1.5cm}{1.3cm}
  \caption*{\changes{$\displaystyle\frac{\omega_r}{\omega_t} = 2\ $, $\omega_t T = 5$}}
\end{subfigure}
  \begin{subfigure}[t]{0.3333\textwidth}
    \def\stackalignment{l}
    \bottominset{(f)}{\includegraphics[width=\linewidth, clip, trim=0.3cm 0.1cm 0.2cm 0.3cm]{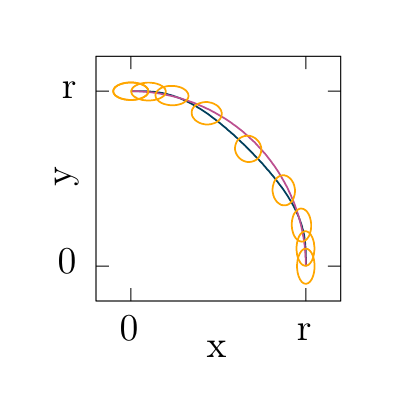}}{1.5cm}{1.3cm}
  \caption*{\changes{$\displaystyle\frac{\omega_r}{\omega_t} = 2\ $, $\omega_t T = 10$}}
\end{subfigure}
  \caption{\changes{
  Trajectories for center of trapping potential (blue) and trap frequencies (yellow ellipses) for shuttling protocols with increasing duration from the left insets to the right insets. The purple line denotes the trajectory of the ion, which traces out the same path in each instance.
  Fast protocols (left) result in substantial deviations between the trajectories of trap center and ion, and this effect is even more pronounced for the more isotropic trapping potentials.}}
  \label{fig:corner-protocol}
\end{figure*}

Fig.~\ref{fig:corner-protocol} depicts some explicit solutions for time-dependent trapping potentials resulting in ground-state to ground-state shuttling protocols.
Insets (a) and (d) correspond to a very fast shuttling protocol with $\omega_tT= 3$,
whereas insets \changes{(b)} and (e) depict instances of a slower shuttling protocol with $\omega_tT=5$,
and insets \changes{(c)} and (f) corresponds to a shuttling protocol with $\omega_tT=10$ that is close to adiabatic.
The insets (a)-(c) correspond to a highly anisotropic trap with $\omega_r=10\omega_t$,
whereas insets (d)-(f) correspond to a less anisotropic trap with $\omega_r=2\omega_t$.

The arc-shaped trajectory $\traj(t)$ of the ion is depicted in purple, and the center of the trapping potential is depicted in blue.
In the close-to-adiabatic cases (c) and (f) the trajectories of ion and center of trapping potential nearly perfectly coincide,
which is consistent with the fact that the ion remains in the instantaneous ground state of the trapping potential in the adiabatic limit.
Deviations between the two trajectories are clearly discernible for the faster shuttling protocols.
In insets (b) and (e) the ion trajectory remains close to the trajectory of the trapping potential, indicating that this protocol is close to the regime of validity of the adiabatic approximation.
In insets (a) and (d), however, the trajectory of the center of trapping potential shares no similarity with the arc-shaped trajectory of the ion.
The ion will thus be substantially displaced from the center of trapping potential, but the carefully chosen dynamics of the trap center ensures that the ion comes to rest at the final instance of the protocol.

Comparison between the instances of the strongly anisotropic trapping potentials featured in (a)-(c) with the instances of less anisotropic trapping potentials (d)-(f) indicate that the isotropy has only a minor impact on the dynamics of the trapping potential in the slow protocols,
but that anisotropy tends to alter dramatically the trajectory of the center of the trapping potential.

The dynamics of the quadratic component of the trapping potential is represented by yellow ellipses indicating equipotential curves,
and these ellipses are depicted for $t=jT/8$ with $j=0,\hdots,8$.
Even in the fast shuttling protocols, the shape of the trapping potential rotates without discernible rapid or non-monotonic dynamics,
which is in quite some contrast to the seemingly wild trajectories of the center of trapping potential.

\changes{In order to gauge the experimental resources required for the present shuttling protocol, it is instructive to specify the maximum of the electric field $\max_{t \in [0,\duration]} \lVert \Hforce(t)/e \rVert$ that is required to keep the ion on its trajectory.
For an Ytterbium-171 ion, with turning radius $r=\SI{30}{\micro\meter}$ \and a trap frequency $\omega_t= 2\pi \times \SI{1}{\mega\hertz}$
this results in an electric field strength
 \be
 \lVert \vec{E} \rVert =2.10 \times 10^{5}  \SI{}{\volt}\SI{}{\meter}^{-1}
 \ee
 for the cases depicted in insets (a) to (c) of Fig.~\ref{fig:corner-protocol},
 and an electric field strength
\be
\lVert \vec{E} \rVert = 8.36 \times 10^{3}  \SI{}{\volt}\SI{}{\meter}^{-1}
\ee
for the cases depicted in insets (d) to (f).
The electric field strength required in these examples thus depend only on the isotropy of the trapping potential, but not on the duration of the shuttling protocol. 
This seems surprising, but also encouraging since it suggests that increasing the speed of shuttling does not require increasingly strong control fields.}

\section{Construction of the invariant}
\label{sec:invariant}

Since the derivation of the invariant introduced in Sec.~\ref{sec:def:invariant} is rather involved,
it is deferred to this Section, with some technical steps discussed in
Appendices \ref{sec:eom} and \ref{sec:real}.

The goal is not to find an invariant for a given Hamiltonian; instead, the time-dependent invariant is the starting point, and the quest is for a Hamiltonian such that Eqs.~\eqref{eq:eominvtwo}, \eqref{eq:eominvone} and \eqref{eq:eominvo} are satisfied for the given invariant $\Inv(t)$.
This can be done in subsequent steps, first determining the quadratic component of the invariant, parametrized in terms of $\Itwo$, and subsequently determining the linear component of the invariant, parametrized in terms of $\Ione$ in Eq.~\eqref{eq:Inv}.

\subsection{A suitable invariant -- the quadratic component}
\label{sec:quadratic}

As stated in Eq.~\eqref{eq:Wansatz},
$\Htwo$ must take the form
\be
\Htwo=\left[
\ba{cc}
m \Htrapfreq & \mathbbm{O}\\
\mathbbm{O} & \frac{1}{m}\um 
\ea\right]\ ,
\ee
where $\Htrapfreq$ is a real symmetric matrix.
The lower right sub block $\frac{1}{m}\um$ is associated with the kinetic energy while the term $\Htrapfreq$ is associated with the potential energy, and is the unknown to be deduced from any proposed invariant.

For a general matrix $\Itwo$, the resultant quadratic part $\Htwo$ of the system Hamiltonian will not be of the desired form given in Eq.~\eqref{eq:Wansatz}, but the parametrization
\be
\Itwo=\Re\left[
\ba{cc}
m \dot\ansatzparam^\dagger\dot\ansatzparam & -\dot\ansatzparam^\dagger\ansatzparam\\
-\ansatzparam^\dagger\dot\ansatzparam & \frac{1}{m} \ansatzparam^\dagger\ansatzparam
\ea\right]\ ,
\label{eq:Itwo}
\ee
with four sub-blocks defined in terms of a $d\times d$-dimensional, {\em complex} matrix $\ansatzparam$ will result in the desired form.
This can be seen by expressing the equation of motion for $\Itwo$ (Eq.~\eqref{eq:eominvtwo}) in terms of $\ansatzparam$,
resulting in
\be \label{eqn:gamma-ansatz}
\left[
\ba{cc}
m \Re(\dot\ansatzparam^\dagger{\cal D}+{\cal D}^\dagger\dot\ansatzparam) & -\Re({\cal D}^\dagger\ansatzparam)\\
-\Re(\ansatzparam^\dagger{\cal D}) & 0
\ea\right]=0\
\ee
where ${\cal D} = \ddot{\ansatzparam} + \ansatzparam \Htrapfreq$.
Any choice of $\ansatzparam$ satisfying the differential equation
\be 
\ddot\ansatzparam+\ansatzparam\Htrapfreq=0\, \label{eqn:rho-eom}
\ee
thus results in a quadratic term $\Itwo$ for the invariant $\Inv$ that is consistent with the initial ansatz in Eq.~\eqref{eq:Wansatz}.

With the current parametrisation of $\Itwo$ in terms of $\ansatzparam$,
one could construct 
$\Htrapfreq$ as
\be
\Htrapfreq=-\ansatzparam^{-1}\ddot\ansatzparam\ .
\ee
Usually, however, this does not result in the identification of a suitable matrix $\Htrapfreq$,
because $\Htrapfreq$ defined in this fashion would not necessarily be symmetric or real.
It is thus necessary to find additional restrictions on $\ansatzparam$ that guarantee the symmetry and reality of $\Htrapfreq$.

To this end, it is helpful to express $\ansatzparam$ in terms of its polar decomposition
\be
\ansatzparam=\unitary\pos\ ,
\label{eq:polar}
\ee
where $\unitary$ is unitary and $\pos$ is Hermitian and positive semi-definite.
It is possible to leave the factor $\pos$ as a free parameter and determine $\unitary$ through the requirement that $\Htrapfreq$ be real-symmetric.

In practice, this task can be formulated in terms of the anti-Hermitian operator
\be
\com=\unitary^\dagger\dot \unitary\ .
\ee
Since $\com$ satisfies the differential equation $\dot \unitary=\unitary\com$, the unitary factor $\unitary$ is uniquely determined in terms of $\com(t)$ and the initial condition $\unitary(0)=\um$.

As derived explicitly in Sec.\ref{sec:eom}, requiring $\Htrapfreq$ to be Hermitian requires $\com$ to be of the form
\be
\com=i\pos^{-2}+\frac{1}{2}[\pos^{-1},\dot \pos]+\frac{1}{2}\pos^{-1}{\antisyminteg}\pos^{-1}\ ,
\label{eq:A}
\ee
with
\be
{\antisyminteg}=\int_{0}^{t}d\tau\ [\Htrapfreq(\tau),\pos(\tau)^2]\ .
\label{eq:J}
\ee
With this specification of $\com$ in terms of $\dot \pos$, $\pos$ and $\Htrapfreq$,
one obtains the differential equation
\be
\{\ddot \pos, \pos\}+\{\pos^2,\Htrapfreq\}=2[\dot\pos,\pos]_{\com}-2\pos\com^2\pos
\label{eq:eomR}
\ee
for $\pos$.

So far, it is only imposed that $\Htrapfreq$ ought to be Hermitian, and not that it should be real as well.
As shown in Sec.~\ref{sec:real}, however, $\Htrapfreq$ is indeed real-symmetric for all times, if $\pos$ is chosen such that it is real and positive at all times and that
\begin{equation}
\dot{\pos}(0)=0\ .
\label{eqn:technical-restriction}
\end{equation}

Even though Eq.\eqref{eq:eomR}, together with Eqs.\eqref{eq:A} and \eqref{eq:J} may be used to determine $\Htrapfreq(t)$,
they are unwieldy in practice, since they relate $\Htrapfreq(t)$ to the integral ${\antisyminteg}$ that depends on the entire history of $\Htrapfreq$.
Since $\Htrapfreq$ is real-symmetric, however, one can -- as discussed in detail in Appendix \ref{sec:eom} arriving at Eq.~\eqref{eqn:j-tilde-equation} -- specify ${\antisyminteg}$ explicitly as
\be
{\antisyminteg}=\sum_{ij}
\frac{(\lambda_i-\lambda_j)(\lambda_i+\lambda_j)^2}{\lambda_i^2+\lambda_j^2}
\matel{\Phi_i}{\dot \pos}{\Phi_j}\ket{\Phi_i}\bra{\Phi_j}
\ee
in terms of the eigenvalues $\lambda_i$ and eigenvectors $\ket{\Phi_i}$ of $\pos$.
With this explicit form of ${\antisyminteg}$, in terms of $\pos$ and $\dot \pos$, but without dependence on $\Htrapfreq$, one can finally solve Eq.~\eqref{eq:eomR} for $\Htrapfreq$, which yields
\be
\Htrapfreq=\sum_{ij}\frac{\matel{\Phi_i}{B}{\Phi_j}}{\lambda_i^2+\lambda_j^2}\ket{\Phi_i}\bra{\Phi_j}\ ,
\label{eq:Omegax}
\ee
with
\be
B=2[\dot\pos,\pos]_\com-2\pos\com^2\pos-\{\ddot \pos, \pos\}\ .
\ee

Any real positive matrix $\pos$ that satisfies the initial condition $\dot\pos(0)=0$ (Eq.~\eqref{eqn:technical-restriction}) thus yields the quadratic term $\Itwo$ of the invariant via Eq.~\eqref{eq:Itwo} that is consistent with the desired form for $\Htwo$ specified in Eq.~\eqref{eq:Wansatz} with a real-symmetric potential matrix $\Htrapfreq$.

Substituting for $\ansatzparam=\unitary \pos$ and $\com=\unitary^\dagger\dot \unitary$ in the Ansatz for $\Gamma$ (Eq.~\eqref{eq:Itwo}),
results in the form of $\Itwo$ that is quoted in Eq.~\eqref{eqn:ansatz-gamma} in Sec.~\ref{sec:def:invariant}. 

\subsection{A suitable invariant -- the linear component}

The goal of this section is to show that the ansatz
\be
\Ione=-\Itwo\left[\ba{c}\traj\\m \dot \traj\ea\right]\ ,
\ee
of Eq.\eqref{eqn:ansatz-w}
results in a linear part $\Hone$ of the Hamiltonian that is of the form
\be
\Hone=\left[\ba{c}-\Hforce \\ 0\ea \right]\ ,
\ee
specified in Eq.~\eqref{eqn:ansatz-v}.

$\Hone$ can be expressed in terms of $\Ione$ as
\be
\Hone={\symplectic}^{-1}\Itwo^{-1}\left(\Htwo{\symplectic}\Ione-\dot \Ione\right)
\ee
via the equation of motion for $\Ione$ given in Eq.~\eqref{eq:eominvone}.
Rather than expressing the right-hand-side in terms of $\Ione$, it will turn out helpful to express it
in terms of $\linearinv$ defined via the relation $\Ione=-\Itwo \linearinv$.
Together with the equation of motion for $\Itwo$ given in Eq.~\eqref{eq:eominvtwo}, this yields
\be
\Hone= {\symplectic}^{-1}\dot \linearinv\ - \Htwo \linearinv.
\ee

With the explicit form of the quadratic term $\Htwo$ of the Hamiltonian specified in Eq.~\eqref{eq:Wansatz} 
this results in the explicit form
\be
\Hone
=
\left[\ba{c}-m \Htrapfreq \linearinv_x - \dot \linearinv_p\\\dot \linearinv_x-\frac{1}{m}\linearinv_p\ea\right]
\ee
for $\Hone$, where the vector $\linearinv$ is decomposed into spatial and momentum parts
\be
\linearinv=\left[
\ba{c}\linearinv_x\\\linearinv_p\ea\right]\ .
\ee 
Finally, choosing $\linearinv_p=m\dot \linearinv_x$ ensures that $\Hone$ is of the form given in Eq.~\eqref{eqn:ansatz-v} as desired,
and the identification of $\traj$ with $\linearinv_x$ results in the sought-after form of $\Ione$ (Eq.~\eqref{eqn:ansatz-w}).

The resulting equation of motion
\be
\ddot \traj + \Htrapfreq \traj = \frac{\Hforce}{m}, \label{eqn:vector-newton-equation}
\ee
permits the determination of $\Hforce$ in terms of $\traj$ and $\pos$ (through $\Htrapfreq$).

\subsection{A suitable invariant -- the scalar component}

The scalar part $\IO(t)$ of the invariant $\Inv$ is determined by Eq.~\eqref{eq:eominvo}.
Strictly speaking, it does not need to be constructed, since the eigenstates of the invariant that are the relevant objects are independent of this phase term.
For the sake of completeness, however, the solution
\be
\IO(t)=\frac{1}{2}\linearinv^T\Itwo \linearinv=-\frac{1}{2}\linearinv^T\Ione=\frac{1}{2}\Ione^T\Itwo^{-1}\Ione\ ,
\ee
which satisfies Eq.~\eqref{eq:eominvo} is stated.

\section{The one-dimensional case}
\label{sec:one-dimensional}

The invariant derived in Sec.~\ref{sec:invariant} does not seem to share many similarities with the invariant that are routinely used in the one-dimensional case~\cite{lewis_direct_1982}.
The regular one-dimensional framework is, however, naturally contained as special case in the present framework, as we will show in the following.

Since all matrices such as $\Htrapfreq$ and $\pos$ reduce to scalars $\Htrapfreq_s$ and $\pos_s$ in the one-dimensional case,
the commutators and generalized commutators vanish in the one-dimensional case.
In particular, the object $\antisyminteg$ defined in terms of a commutator in Eq.~\eqref{eq:J} vanishes,
so that Eq.~\eqref{eq:A} simplifies to
\be \label{eqn:one-d-a}
\com_s=i\pos_s^{-2}.
\ee
Substituting Eq.~\eqref{eqn:one-d-a} into Eq.~\eqref{eq:eomR} and exploiting the fact that all objects commute, gives that
\be \label{eqn:pre-ermakov-equation}
\ddot{\pos}_s \pos_s + \Htrapfreq_s \pos_s^2  = \pos_s^{-2}\ .
\ee

The quadratic part $\Gamma$ defined in Eq.~\eqref{eqn:ansatz-gamma} simplifies to
\bqa
\Itwo_s=\left[
  \ba{cc}
  \displaystyle m \left( \dot{\pos}_s^2 + \pos_s^{-2} \right) &  
  \displaystyle - \pos_s \dot{\pos}_s \\
   \displaystyle  - \pos_s \dot{\pos}_s &
  \frac{1}{m} \pos_s^2
  \ea \right]\ ,
\label{eqn:simplified-gamma}
\eqa
in the case of commutativity.
From Eqs.~\eqref{eqn:ansatz-w}, \eqref{eqn:ansatz-scalar} and \eqref{eqn:simplified-gamma}, it is possible to write the invariant $\Inv$ out explicitly in terms of $\hat{r}$ and $\hat{p}$,
\begin{align}
\Inv_s
&= \frac{1}{2m} \left(\pos_s (\hat{p} - m \dot{L}) - m \dot{\pos}_s (\hat{r} - L) \right)^2 \nonumber \\
&+ \frac{1}{2}m \left( \frac{\hat{r} - L}{\pos_s} \right)^2, 
\label{eqn:one-d-invariant}
\end{align}
which is the Ermakov-Lewis invariant \cite{lewis_classical_1967} with additional terms linear in position and momentum \cite{lewis_direct_1982}.
Dividing both sides of Eq.~\eqref{eqn:pre-ermakov-equation} by $\pos_s$ gives
\be \label{eqn:one-d-ermakov}
\ddot{\pos}_s + \pos_s \Htrapfreq_s = \pos_s^{-3},
\ee
which is the Ermakov equation that appears in the original treatment \cite{lewis_classical_1967},
where the symbol $\rho$ is used instead of $\pos_s$,
and the symbols $\omega^2$ and $\alpha$  are used instead of $\Htrapfreq_s$ and $\traj$.

\section{Outlook}

The extension of the framework of quantum invariants to systems with quadratic potentials of more than one translational degree of freedom
opens up a wide perspective for the control of quantum systems.
While the present work is motivated by control of trapped ions, none of the features of the present framework are specific to this system,
and the only precondition for the use of the presented invariant is the validity of the Gaussian approximation.
\changes{Since also shuttling of cold neutral atoms has been experimentally realised in one-dimensional optical lattice potentials~\cite{e22030262} with time-dependent potential designed with the Ermakov-Lewis invariant, the presently derived framework shall open up possibilities to extend such tasks to higher-dimensional optical lattices.}

The ability to systematically devise time-dependent potentials that realise ground-state to ground-state transfer permits to identify those potentials that are optimized with respect to additional desirable properties.
Typical examples might include shuttling protocols with particularly weak forcing or potentials that suit best the geometry of the electrodes that generate the potentials.
As such, the present framework promises to substantially advance the shuttling of trapped ions, which is a central step towards achieving scalability of quantum information technology with trapped ions.

The specific choice of invariant discussed here, does not necessarily need to be seen as the unique generalization of one-dimensional invariants,
but rather as evidence that such generalizations are possible.
It thus seems conceivable that different generalization can be found, and that different restrictions of the achievable Hamiltonians can be found.
One may, for example, envision the Lorentz force caused by a homogeneous magnetic field and aim at finding an invariant that ensures that no inhomogeneities in the magnetic field arise.

In addition to the extension of experimental tasks such as noise-resilient shuttling \cite{levy_noise_2018,lu_fast_2014,lu_fast_2018} to higher dimensional settings,
the present work can also initiate new steps in the conceptual developments of quantum invariants.

\section*{Acknowledgements}

We are indebted to stimulating discussions with Adam Callison, Alexander Paige, Pedro Taylor-Burdett, Sebastian Weidt and Winnie Hensinger.
\changes{This work was supported through a studentship in the
Centre for Doctoral Training on Controlled Quantum Dynamics at Imperial College London funded by EPSRC(EP/L016524/1).}

\appendix

\section{Derivation of the invariant equations}
\label{sec:eom}

This section contains the explicit derivation of Eqs.~\eqref{eq:A},\eqref{eq:J} and \eqref{eq:eomR}.

The equation of motion $\ddot\ansatzparam+\ansatzparam\Htrapfreq=0$ (Eq.~\eqref{eqn:rho-eom})
with $\ansatzparam$ expressed in terms of its polar decomposition $\ansatzparam=\unitary\pos$ (Eq.~\eqref{eq:polar}) results in
\begin{equation} \label{eqn:first-step}
K:=\dot \com \pos + \com^2 \pos + 2 \com \dot{\pos} + \ddot{\pos} + \pos\Htrapfreq=0\ .
\end{equation}
Since $M$ is Hermitian, and the relation $\pos K=K^\dagger \pos$ holds because $K$ vanishes,
this is equivalent to
\begin{align}
\frac{d}{dt}\pos \com \pos = &
\frac{1}{2}\left([\ddot \pos, \pos]   +[ \Htrapfreq, \pos^2]\right)\ .
 \label{eqn:antisymmetric-part}
\end{align}
The right hand side of Eq.~\eqref{eqn:antisymmetric-part} can be rewritten in terms of a total derivative to obtain
\begin{equation}
\frac{d}{dt}  \pos\com\pos = \frac{1}{2} \frac{d}{dt}\left( [\dot \pos, \pos] + {\antisyminteg} \right),
\end{equation}
with
\be
{\antisyminteg} = \int^t _0 d \tau\ [\Htrapfreq(\tau),\pos(\tau)^2]\ .
\ee
This can be directly integrated, resulting in the explicit solution
\begin{equation}
\com = \pos^{-1} C \pos^{-1} + \frac{1}{2}[\pos^{-1}, \dot{\pos} ]+ \frac{1}{2} \pos^{-1} {\antisyminteg} \pos^{-1},
\end{equation}
for $\com$.

The matrix $C$ is the constant of integration that can be chosen freely.
With the choice $C=i \um$ motivated by the Ermakov equation in the one-dimensional case \cite{lewis_classical_1967,leach_ermakov_2008},
this results in Eq.~\eqref{eq:A}.

Using again the fact that $K$ vanishes, we can form the relation $\pos K + K^\dagger \pos = 0$ and rearrange to derive Eq.~\eqref{eq:eomR}.

\section{Reality of $\Htrapfreq$}
\label{sec:real}

This section contains the proof that $\Htrapfreq(t)$ determined by Eq.\eqref{eq:eomR} is indeed real and symmetric for all times, provided that the initial condition $\dot \pos(0)=0$ is satisfied.

Eq.~\eqref{eq:eomR} reads explicitly
\be \label{eqn:first-equation}
\{\ddot{\pos}, \pos \} + \{\pos^2, \Htrapfreq\} = 2 [\dot{\pos},\pos]_{\com}  - 2 \pos \com^2 \pos\ ,
\ee
and there is indeed no evident reason why $\Htrapfreq$ determined from this relation together with Eqs.~\eqref{eq:A} and \eqref{eq:J} should be real and symmetric.

In the following, explicit, closed-form expressions for ${\antisyminteg}$, $\com$ and $\Htrapfreq$ are derived, that hold under the assumption that $\Htrapfreq$ is real and symmetric.
It is then demonstrated that such closed-form expressions form indeed a solution of Eqs.~\eqref{eq:A}, \eqref{eq:J} and ~\eqref{eq:eomR}, which not only completes the proof that $\Htrapfreq$ is real symmetric, but gives a direct way to calculate $\Htrapfreq$.

\subsection{Necessary conditions for the reality of $\Htrapfreq$}

Under the assumption that $\Htrapfreq$ is real and symmetric,
the imaginary part of the equation of motion for $\pos$ (Eq.~\eqref{eq:eomR}) reads
\be
\{{\antisyminteg},\pos^{-2}\} = [\dot \pos,\pos^{-1} ]+ [\pos,\pos^{-2}]_{\dot\pos}\ .
\label{eqn:j-definition}
\ee
This determines ${\antisyminteg}$ in terms of $\pos$ and $\dot \pos$, and, in contrast to the defining relation (Eq.~\eqref{eq:J}) for ${\antisyminteg}$ it is time-local.

In order to solve Eq.~\eqref{eqn:j-definition} for ${\antisyminteg}$, it is helpful to consider the eigen-decomposition
\be
\pos=\sum_{j}\lambda_j\prj{\Phi_j}
\ee
of the real and positive semi-definite matrix $\pos$.\\
In terms of matrix elements with respect to the eigenstates $\ket{\Phi_j}$ of $\pos$,
Eq.~\eqref{eqn:j-definition} can directly be solved, yielding
\begin{equation} \label{eqn:j-tilde-equation}
{\antisyminteg}_{jk} = \frac{(\lambda_j - \lambda_k) (\lambda_j + \lambda_k)^2}{\lambda_j ^2 + \lambda_k ^2}{\dot{\pos}}_{jk}\ ,
\end{equation}
where
\be
{O}_{jk}=\matel{\Phi_j}{O}{\Phi_k}
\ee
is a short hand notation for the matrix elements of operator $O$,
and
\be
\dot O_{jk}=\matel{\Phi_j}{\dot O}{\Phi_k}
\ee
does generally {\em not} coincide with $\frac{\partial}{\partial t}O_{jk}$ because of the time-dependence of the eigenstates $\ket{\Phi_j}$.

Similarly to Eq.\eqref{eqn:j-tilde-equation}, also Eq.~\eqref{eq:eomR} can be solved for $\Htrapfreq_{jk}$, resulting in
\begin{widetext}
\be
\label{eqn:m-tilde-definition}
{\Htrapfreq}_{jk}=
\frac{1}{\lambda_{j}^2 + \lambda_{k}^2}
\left(
2\sum_l\left(\lambda_k{\dot{\pos}}_{jl} A_{lk} -\lambda_jA_{jl}\dot{\pos}_{lk}-\lambda_j\lambda_kA_{jl}A_{lk}\right)-{\ddot{\pos}}_{jk} \left(\lambda_j + \lambda_k \right)
\right)\ .
\ee
Eq.~\eqref{eq:A} expressed similarly in terms of matrix elements reads
\be
\label{eqn:a-tilde-definition}
{\com}_{jk} = \frac{1}{2 \lambda_j  \lambda_k }\left(
2i\delta_{jk}-(\lambda_j - \lambda_k){\dot{\pos}}_{jk}+{\antisyminteg}_{jk}\right)\ .
\ee
Substituting Eq.~\eqref{eqn:j-tilde-equation} into Eq.~\eqref{eqn:a-tilde-definition} yields
\begin{equation} \label{eqn:a-tilde-equation}
  \com_{jk}=\frac{i }{\lambda_j \lambda_k} \delta_{jk} + \frac{\lambda_j - \lambda_k}{\lambda_j ^2 + \lambda_k ^2} \dot{R}_{jk}.
\end{equation}
Substituting Eq.~\eqref{eqn:a-tilde-equation} into Eq.~\eqref{eqn:m-tilde-definition} yields
\begin{equation} \label{eqn:m-tilde-equation}
{\Htrapfreq}_{jk}=
 \frac{\delta_{jk}}{\lambda_j ^4} - \frac{(\lambda_j + \lambda_k)}{\lambda_j ^2 + \lambda_k ^2}{\ddot{\pos}}_{jk} +
 2\sum_l {\dot{\pos}}_{jl}{\dot{\pos}}_{lk} p(\lambda_j, \lambda_k, \lambda_l)\ ,
\end{equation}
with
\be
p(\lambda_j, \lambda_k, \lambda_l) =
\frac{
\lambda_l^3\left(\lambda_j+\lambda_k\right)-
\lambda_l^2\left(\lambda_j ^2+\lambda_k^2- \lambda_j \lambda_k\right)-
\lambda_j ^2 \lambda_k ^2}
{(\lambda_j ^2 + \lambda_k ^2) (\lambda_j^2 + \lambda_l^2)(\lambda_k ^2 + \lambda_l ^2)}.
\ee
\end{widetext}

The closed-form expressions for
${\antisyminteg}$, $\com$ and ultimately $\Htrapfreq$, in terms of their matrix elements, given in Eqs,~\eqref{eqn:j-tilde-equation}, \eqref{eqn:a-tilde-equation} and \eqref{eqn:m-tilde-equation}, hold if $\Htrapfreq$ is real symmetric.

Conversely, if Eqs.~\eqref{eqn:j-tilde-equation}, \eqref{eqn:a-tilde-equation} and \eqref{eqn:m-tilde-equation} hold, then $\Htrapfreq$ is guaranteed to be real symmetric since Eq.~\eqref{eqn:m-tilde-equation} is manifestly symmetric and defined solely using real quantities.
As a consequence, $\Htrapfreq$ is real symmetric if and only if these expressions hold.

\subsection{Direct proof of the reality of $\Htrapfreq$}

What remains is thus to verify directly that taking Eqs.~\eqref{eqn:j-tilde-equation}, \eqref{eqn:a-tilde-equation} and \eqref{eqn:m-tilde-equation} as defining relations for $\antisyminteg$, $\com$ and $\Htrapfreq$, implies that Eqs.~\eqref{eq:A}, \eqref{eq:J} and \eqref{eq:eomR} are satisfied.
Doing this completes the proof that $\Htrapfreq$ is real symmetric.

Henceforth it is assumed that Eqs.~\eqref{eqn:j-tilde-equation}, \eqref{eqn:a-tilde-equation} and \eqref{eqn:m-tilde-equation} hold.
As Eq.~\eqref{eqn:a-tilde-equation} is derived using Eq.~\eqref{eqn:a-tilde-definition} which is Eq.~\eqref{eq:A} expressed in terms of the matrix elements of the eigenstates of $\pos$, Eq.~\eqref{eq:A} holds by construction.
Similarly, Eq.~\eqref{eqn:m-tilde-equation} is derived from Eq.~\eqref{eqn:m-tilde-definition}, which itself is Eq.~\eqref{eq:eomR} expressed in terms of matrix elements, so Eq.~\eqref{eq:eomR} additionally holds.

It remains to show that Eq.~\eqref{eq:J},
or its equivalent differential form
\begin{align} \label{eqn:differential-j-form}
{\antisyminteg}(0)&=0\ ,\ \mbox{and} \\
\dot{\antisyminteg}&= \pos^2 \Htrapfreq - \Htrapfreq \pos^2\ . \label{eqn:second-differential-j-form}
\end{align}
is satisfied
Since $\dot{\pos}(0)=0$, Eq.~\eqref{eqn:differential-j-form} follows directly from inspection of Eq.~\eqref{eqn:j-tilde-equation}.

Eq.\eqref{eqn:second-differential-j-form} reads in terms of matrix elements
\be \label{eqn:j-form-matrix-elements}
\dot{\antisyminteg}_{jk} = (\lambda_j ^2 - \lambda_k ^2) \Htrapfreq_{jk}.
\ee
In order to complete the proof, it is sufficient to show that Eq.~\eqref{eqn:j-form-matrix-elements} is satisfied. As $\antisyminteg$ is defined in terms of its matrix elements $\antisyminteg_{jk}$ in Eq.~\eqref{eqn:j-tilde-equation}, it is necessary to relate these to the $\dot{\antisyminteg_{jk}}$ that appear in Eq.~\eqref{eqn:j-form-matrix-elements}.

The time-derivative of $\antisyminteg_{jk}$ reads
\be \label{eqn:j-curvature}
\frac{d}{dt} \antisyminteg_{jk} = \sum_l W_{jl} \antisyminteg_{lk} - \sum_l \antisyminteg_{jl} W_{lk} + \dot{\antisyminteg}_{jk}
\ee
with
\be
W_{jk}=\langle{{\dot\Phi}_j}|{\Phi_k}\rangle.
\ee

Substituting this expression into Eq.~\eqref{eqn:j-form-matrix-elements} gives
\begin{equation}
\frac{d}{dt}{{{\antisyminteg}}}_{jk} = \sum_l W_{jl} {\antisyminteg}_{lk} - \sum_l {\antisyminteg}_{jl} W_{lk} + (\lambda_j^2 - \lambda_k^2){\Htrapfreq}_{jk}, \label{eq:equiv-j}
\end{equation}
which is equivalent to Eq.~\eqref{eqn:j-form-matrix-elements}.

The next step is to differentiate Eq.~\eqref{eqn:j-tilde-equation} and use the result to show that Eq.~\eqref{eq:equiv-j} is satisfied identically.

Defining, motivated by Eq.~\eqref{eq:equiv-j},
\begin{equation} \label{eqn:xi-definition}
\Xi = \frac{d}{dt}{\antisyminteg}_{jk} - \sum_l W_{jl} {\antisyminteg}_{lk} + \sum_l {\antisyminteg}_{jl} W_{lk} + (\lambda_k^2 - \lambda_j^2) {\Htrapfreq}_{jk},
\end{equation}
the remaining task is to show that $\Xi=0$.

Substituting for ${\antisyminteg}_{jk}$ and ${\Htrapfreq}_{jk}$ using Eqs.~\eqref{eqn:j-tilde-equation} and ~\eqref{eqn:m-tilde-equation} in Eq.~\eqref{eqn:xi-definition}
gives
\begin{widetext}
\begin{align} \label{eqn:big-mess}
\Xi &=
\frac{d}{dt}\left( \frac{(\lambda_j ^2 - \lambda_k ^2)(\lambda_j + \lambda_k )}{\lambda_j ^2 + \lambda_k ^2} \right) {\dot{\pos}}_{jk}+
\frac{(\lambda_j ^2 - \lambda_k ^2)(\lambda_j + \lambda_k )}{\lambda_j ^2 + \lambda_k^2} \frac{d}{dt}\left({{\dot{\pos}}}_{jk}\right) \nonumber \\
&- \sum_l \frac{(\lambda_l ^2 - \lambda_k ^2)(\lambda_l + \lambda_k)}{\lambda_l ^2 + \lambda_k ^2} W_{jl}{\dot{\pos}}_{lk}
+ \sum_l \frac{(\lambda_j ^2 - \lambda_l ^2)(\lambda_j + \lambda_l)}{\lambda_j ^2 + \lambda_l ^2} {\dot{\pos}}_{jl}W_{lk} \nonumber \\
&+ \frac{ (\lambda_j + \lambda_k) (\lambda_k ^2 - \lambda_j ^2) }{\lambda_j ^2 + \lambda_k ^2} {\ddot{\pos}}_{jk}
- 2 \left( \lambda_k^2 - \lambda_j^2 \right)  p(\lambda_j, \lambda_k, \lambda_l) {\dot{\pos}}_{jl} {\dot{\pos}}_{lk}\ .
\end{align}

Given that the states $\ket{\Phi_i}$ are the eigenvectors of $\pos$,
the explicit expressions for $\pos_{ij}$, $\dot{\pos}_{ij}$ and $\ddot{\pos}_{ij}$ read
\begin{align}
R_{jk} &= \lambda_j \delta_{jk}\ ,\\
\dot{\pos}_{jk} &= ( \lambda_j - \lambda_k) W_{jk} + \dot{\lambda}_j \delta_{jk}\ ,\label{eqn:prelim-identity-1}\\
{\ddot{\pos}}_{jk} &= \sum_l (\lambda_j + \lambda_k - 2  \lambda_l) W_{jl} W_{lk} + (\lambda_j - \lambda_k) \dot{W}_{jk} 
+ 2 (\dot{\lambda}_j - \dot{\lambda}_k) W_{jk} + \ddot{\lambda}_j \delta_{jk}\ . \label{eqn:prelim-identity-2}
\end{align}

Substituting Eq.~\eqref{eqn:prelim-identity-1} and Eq.~\eqref{eqn:prelim-identity-2} into Eq.~\eqref{eqn:big-mess} gives
\begin{equation}
\Xi = W_{jk} r(\lambda_j, \lambda_k, \dot{\lambda}_j , \dot{\lambda}_k) + \sum_l W_{jl}W_{lk} s(\lambda_j,\lambda_k,\lambda_l),
\end{equation}
where
\begin{align}
r(\lambda_j, \lambda_k, \dot{\lambda}_j , \dot{\lambda}_k)&= (\lambda_j - \lambda_k) \frac{d}{dt} \left( \frac{(\lambda_j ^2 - \lambda_k ^2) (\lambda_j + \lambda _k)}{\lambda_j ^2 + \lambda_k ^2}\right)
- \frac{(\lambda_j ^2 - \lambda _k ^2)(\lambda_j + \lambda_k)(\dot{\lambda}_j - \dot{\lambda}_k)}{\lambda_j ^2 + \lambda_k ^2} \nonumber \\
  & - 2 (\lambda_j - \lambda_k) (\lambda_k ^2 - \lambda_j ^2)
\left(\dot{\lambda}_j p(\lambda_j, \lambda_k, \lambda_j) + \dot{\lambda}_k p(\lambda_j, \lambda_k, \lambda_k)\right)\ ,
\end{align}
and 
\begin{align}
s(\lambda_j, \lambda_k, \lambda_l) &=
-\frac{(\lambda_l ^2 - \lambda_k ^2)^2}{\lambda_l^2 + \lambda_k^2} + \frac{(\lambda_j ^2 - \lambda_l ^2)^2}{\lambda_j^2 + \lambda_l^2}
 + \frac{(\lambda_j + \lambda_k)(\lambda_k^2  - \lambda_j^2)(\lambda_j + \lambda_k - 2\lambda_l)}{\lambda_j^2 + \lambda_k^2} \nonumber \\
  &- 2(\lambda_j - \lambda_l)(\lambda_k - \lambda_l)(\lambda_j ^2 - \lambda_k ^2) p(\lambda_j, \lambda_k, \lambda_l)\ .
\end{align}
Both $r$ and $s$ are rational functions of their arguments, and they are indeed identically vanishing. 
Therefore $\Xi=0$ which completes the proof.

\end{widetext}

\bibliography{bib_invariants}

\end{document}